\documentclass{eas}
\usepackage{graphicx}
\usepackage{graphics}
\usepackage{amsmath}
\usepackage{amssymb}
\usepackage{gensymb}
\usepackage[version=4]{mhchem}
\usepackage{subfigure}

\TitreGlobal{Conditions and Impact of Star Formation}
\begin{document}

%%-----------------------------
%%      the top matter
%%-----------------------------

\title{Chemical complexity in protoplanetary disks in the era of ALMA and Rosetta} 
\author{Catherine Walsh}
\address{Leiden Observatory, Leiden University, P.~O.~Box 9513, 2300 RA Leiden, The Netherlands}

\runningtitle{Chemical complexity in protoplanetary disks}

\begin{abstract}
Comets provide a unique insight into the molecular composition and complexity of the material 
in the primordial solar nebula.  
Recent results from the {\em Rosetta} mission, currently monitoring comet 67P/Churyumov-Gerasimenko {\em in situ}, 
and ALMA (the Atacama Large Millimeter/submillimeter Array) have demonstrated a tantalising link between 
the chemical complexity now confirmed in disks 
(via the detection of gas-phase \ce{CH3CN}; \"{O}berg \etal \cite{oberg15}) and that confirmed on the 
surface of 67P (Goesmann \etal \cite{goesmann15}),    
raising questions concerning the chemical origin of such species (cloud or inheritance versus disk synthesis).
Results from an astrochemical model of a protoplanetary disk are presented in which 
complex chemistry is included and in which it is assumed that simple ices only are inherited from the 
parent molecular cloud.     
The model results show good agreement with the abundances of several COMs observed 
on the surface of 67P with {\em Philae}/COSAC. 
Cosmic-ray and X-ray-induced photoprocessing of predominantly simple ices inherited by the 
protoplanetary disk is sufficient to generate a chemical complexity similar to that observed in comets.  
This indicates that the icy COMs detected on the surface of 67P may have a disk origin.
The results also show that gas-phase \ce{CH3CN} is abundant in the inner warm disk 
atmosphere where hot gas-phase chemistry dominates and potentially erases the ice 
chemical signature.  
Hence, \ce{CH3CN} may not be an unambiguous tracer of the complex organic ice reservoir.   
However, a better understanding of the hot gas-phase chemistry of \ce{CH3CN} is needed to confirm this 
preliminary conclusion. 
\end{abstract}

\maketitle

%%-----------------------------
%%      your text
%%-----------------------------

\section{Probing ices in protoplanetary disks}
Protoplanetary disks are the reservoirs of the basic components - dust, gas, and ice - 
required for the formation of planetary systems.
The molecular components of midplane ices, in particular, sets the initial 
composition of icy planetesimals which can coalesce to form 
comets, and/or become swept up by forming planets.  
The chemical heritage of this icy planet- and comet-building material 
is much debated. 
Theories range from inheritance from the parent molecular cloud, chemistry en 
route from protostellar envelope to protoplanetary disk, to chemical processing 
within the protoplanetary disk once formed.  
Related to this is the origin of chemical complexity in 
planetary systems.  
The formation of many saturated (or close to saturated) organic molecules 
(\vg, \ce{CH3OH}) is thought to occur on or within icy mantles on dust grains.  
Such species are considered an important bridge between 
the simple molecules generally detected in space, and 
those considered important for prebiotic chemistry, \vg, amino acids.

It remains challenging to directly observe the icy 
planet-building material in protoplanetary disks (with the exception of 
water ice; \vg, Pontoppidan \etal \cite{pontoppidan05};  
Terada \etal \cite{terada07}; McClure \etal \cite{mcclure15}).  
However, once the ice reservoir is released from the 
dust grains, either via heating (thermal desorption) or triggered 
by the absorption of energetic photons or particles (non-thermal desorption) 
the composition can be indirectly probed by gas-phase observations.  
Several of the expected dominant ice components 
(\vg, \ce{H2O}, \ce{CO2}, \ce{CO}, and \ce{CH4}) have been detected 
in the gas-phase in the warm/hot ($\gtrsim 300$~K, $\lesssim 10$~AU) 
inner regions of protoplanetary disks (\vg, Carr \& Najita \cite{carr08}; 
Gibb \& Horne \cite{gibb13}; Lahuis \etal \cite{lahuis06}; Mandell \etal \cite{mandell12}).  
However, such observations probe the disk upper layers only and may reveal a composition 
different to that in the disk midplane within which the bulk of the planet-building 
reservoir resides (\vg, Walsh \etal \cite{walsh15}).  
The indirect detection of the outer ($\gtrsim10$~AU) ice reservoir is more challenging. 
Here, non-volatile species (\vg, \ce{H2O}) 
are thought to be desorbed non-thermally by UV photons from the 
central star and/or interstellar medium and are predicted to reside in a 
narrow layer bounded above by photodissociation and below by 
freezeout.  
Non-thermal desorption is less efficient than thermal desorption and so the peak 
abundances reached in the outer regions are orders of magnitude less than expected 
in the thermally desorbed regions.  
The first detection of 
cold water vapour in a protoplanetary disk with {\em Herschel} supports this picture 
(Hogerheijde \etal \cite{hogerheijde11}).  
  
COMs (loosely defined as consisting of 6 or more atoms; 
Herbst \& van~Dishoeck \cite{herbst09})~have a similar 
volatility as water and the additional challenge of intrinsically weaker 
emission arising from their inherent molecular complexity and typically lower 
abundances.  
Hence, very high sensitivity observations are required to detect emission from 
such gas-phase species, especially from small 
sources which span a few arcseconds only on the sky (such as protoplanetary disks).  
Luckily, we are now in the era of ALMA which has increased the achievable sensitivity 
of (sub)mm observations by several orders of magnitude.  
Indeed, \"{O}berg \etal (\cite{oberg15}) recently report the 
ALMA detection of a complex molecule, \ce{CH3CN}, 
in a protoplanetary disk (MWC~480) for the first time.    
\"{O}berg \etal (\cite{oberg15}) derive an abundance ratio for \ce{CH3CN}/\ce{HCN} between 
$\approx5$\% and 20\% in line with that detected towards cometary comae 
($\approx10$\%; Mumma \& Charnley \cite{mumma11}) suggesting that the emission 
arises from thermal desorption of the comet-building ice reservoir.  
%The central star, MWC~480, is a Herbig~Ae star ($T_\mathrm{eff}\approx 8,000$~K), 
%which creates a larger thermally-desorbed reservoir in the inner disk, relative 
%to that expected in disks around cooler stars.  

%\footnotetext{COMs are loosely defined as those consisting of 6 or more atoms 
%(\vg, Herbst \& van~Dishoeck \cite{herbst09}).}

In parallel, the \ce{Rosetta} mission is monitoring the chemical 
composition of 67P {\em in situ}.  
The COSAC (COmetary SAmpling and Composition) experiment on board the lander, 
{\em Philae}, has revealed a plethora of sub-surface COMs, 
including \ce{CH3CN}, \ce{NH2CHO}, \ce{CH3COCH3}, and \ce{CH3OCN}, amongst others 
(Goesmann \etal \cite{goesmann15}). 
The \ce{CH3CN}/\ce{HCN} ratio measured with COSAC ($\approx30$\%) 
is very similar to that observed in the disk around MWC~480 (\"{O}berg \etal \cite{oberg15}).

In light of these new quantitative results from ALMA and {\em Rosetta}, I revisit 
the models presented in Walsh \etal (\cite{walsh14}) which compute 
the abundance and distribution of COMs in a protoplanetary disk 
around a T~Tauri star.  
One of the aims of these models was to test whether surface chemistry is able to 
efficiently synthesise COMs in protoplanetary disks given that the 
disk inherits only simple ices 
(\ce{H2O}, \ce{CO2}, \ce{CO}, \ce{N2}, \ce{CH4}, \ce{NH3}, and \ce{CH3OH}).  
In Sect.~\ref{ch3cn} I discuss the abundance and distribution of \ce{CH3CN} gas and ice, and
in Sect.~\ref{icycoms} I discuss the predicted abundances of those icy complex molecules 
detected in the sub-surface layers of 67P.  
I end with a short summary and future outlook (Sect.~\ref{summary}).

%\begin{figure}
%\caption{Cartoon showing the gas reservoirs generated by the desorption of 
%the ice reservoir via thermal desorption (where $T \ge T_\mathrm{des}$, orange regions) and 
%non-thermal desorption by ultraviolet photons in the outer disk (turquoise regions).  
%For some molecules (\vg, \ce{H2O} and \ce{HCN}) there are also hot 
%gas-phase routes to formation in the inner warm disk atmosphere (\vg, Walsh \etal \cite{walsh15}).}
%\subfigure{\includegraphics[width=\textwidth]{./reservoirs.pdf}}
%\label{figure1}
%\end{figure}

\section{\ce{CH3CN} gas in protoplanetary disks}
\label{ch3cn}

\begin{figure}
\caption{Abundance of \ce{CH3CN} gas (orange) and ice (blue) relative 
to \ce{H2} (top row) as a function of disk radius, $R$, and 
height (scaled by the radius, $Z/R$), compared with the same data 
for \ce{CH3OH} gas and ice (bottom row).  Data are from Walsh \etal (\cite{walsh14}).}
\subfigure{\includegraphics[width=\textwidth]{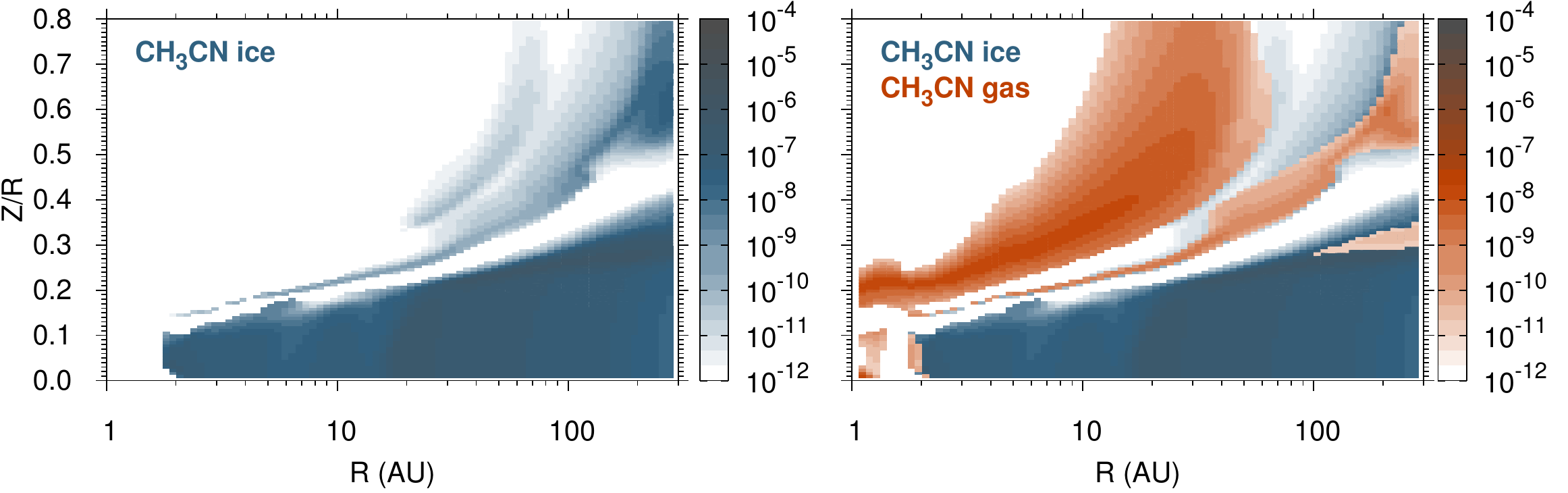}}
\subfigure{\includegraphics[width=\textwidth]{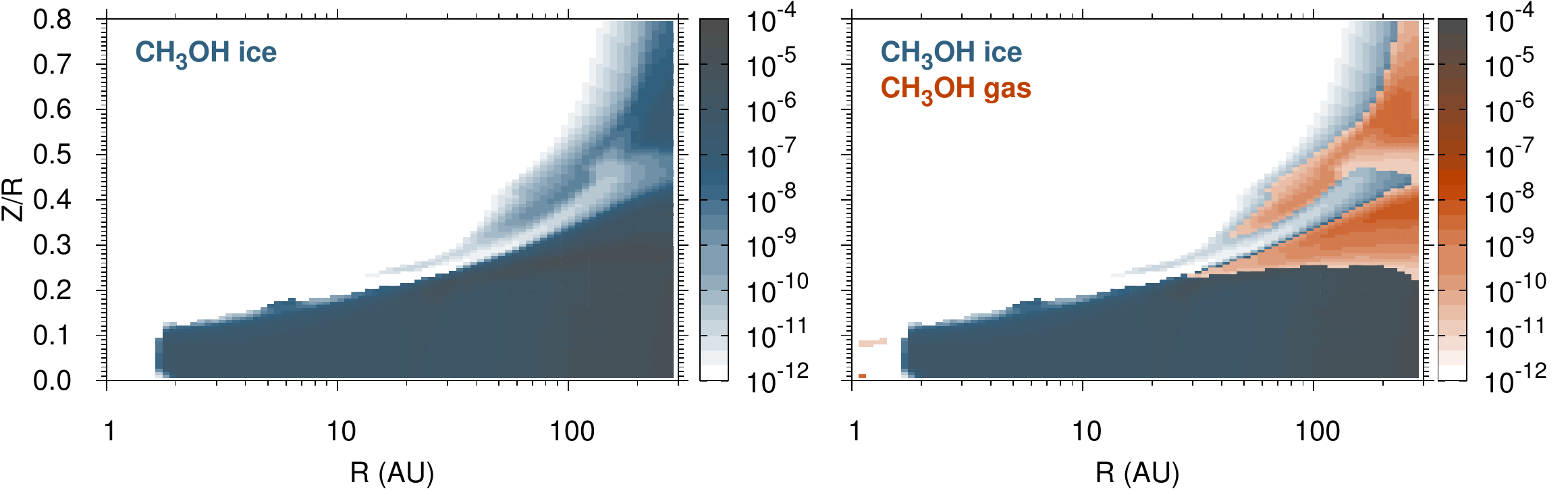}}
\label{figure2}
\end{figure}

In Fig.~\ref{figure2} I display the fractional abundance (relative to \ce{H2}) 
of \ce{CH3CN}  (top row) and \ce{CH3OH} (bottom row) gas (orange) and ice (blue), 
as a function of disk radius and height.  
The physical conditions are those for an irradiated disk in hydrostatic 
equilibrium around a typical T~Tauri star, and the chemical network includes 
gas-phase chemistry, gas-grain interactions, and grain-surface chemistry 
(see Nomura \etal \cite{nomura07} and Walsh \etal \cite{walsh14} for full details).  
\ce{CH3CN} and \ce{CH3OH} ice show a similar distribution with the 
complex ices reaching their peak abundance 
($\sim10^{-6}-10^{-5}$ relative to \ce{H2}, respectively) 
towards the disk midplane and towards the outer regions of the disk.  
As a comparison, water ice has a peak abundance a few times $10^{-4}$ relative to \ce{H2}.
\ce{CH3CN} gas has two reservoirs, a ``hot'' reservoir in the inner disk atmosphere 
within $\approx~30$~AU and a narrow photodesorbed layer in outer disk tracing the 
boundary of the \ce{CH3CN} ice layer.  
In contrast, \ce{CH3OH} gas has an outer photodesorbed layer only.  
In the inner warm disk atmosphere, the chemical abundances are mediated 
by formation via gas-phase chemistry and/or desorption of the ice reservoir 
and destruction by gas-phase chemistry and photodissociation by the stellar 
radiation field.  
The results show that \ce{CH3CN} gas can survive where \ce{CH3OH} gas cannot, 
which potentially points to an efficient gas-phase formation route to \ce{CH3CN}, \ie, 
\ce{CH3CN} is not solely reliant on 
grain-surface synthesis, in contrast with \ce{CH3OH} 
(as also discussed in \"{O}berg \etal \cite{oberg15}).  
An alternative explanation is that the models are lacking destruction mechanisms 
for gas-phase \ce{CH3CN}.  
The results show that \ce{CH3CN} in the inner regions of protoplanetary disks 
may not be directly tracing the composition of the comet-forming zone; 
however, a better understanding of the gas-phase chemistry of \ce{CH3CN} 
in the inner warm regions of protoplanetary disks is required to 
confirm this preliminary conclusion.  

\section{\ce Icy complex molecules in the comet-forming zone}
\label{icycoms}

\begin{figure}
\caption{Percentage of \ce{CH3CN}, \ce{NH2CHO}, \ce{HOCH2CHO}, and 
\ce{CH3COCH3} ice relative to water ice, as a function of disk radius, $R$, and 
height (scaled by the radius, $Z/R$).   
Data are from Walsh \etal (\cite{walsh14}).}
\subfigure{\includegraphics[width=\textwidth]{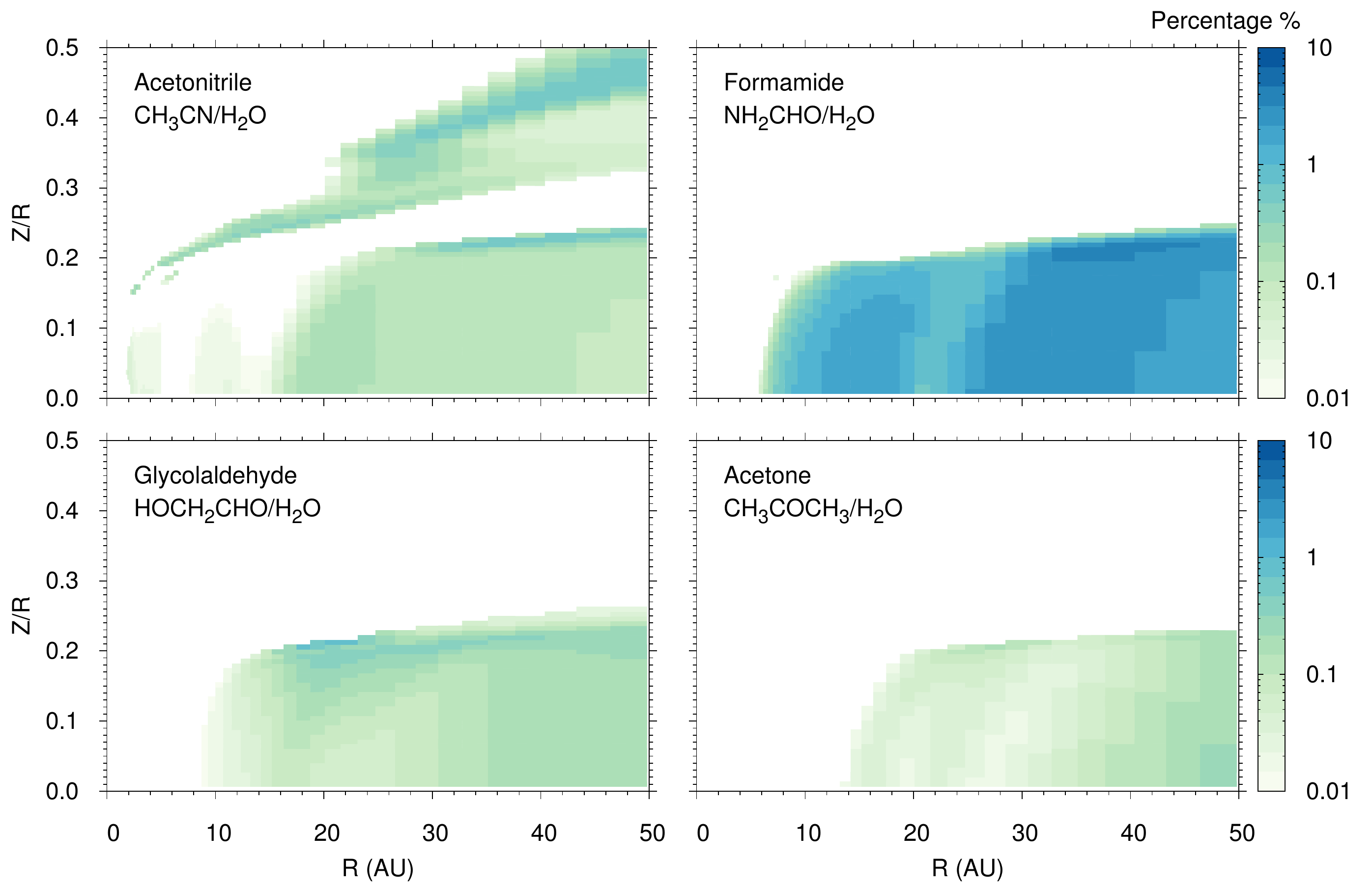}}
\label{figure3}
\end{figure}

In Fig.~\ref{figure3} I display the ice abundances of four complex 
molecules calculated using the same disk model 
(\ce{CH3NH2}, \ce{NH2CHO}, \ce{HOCH2CHO}, and \ce{CH3COCH3}).  
These species have all been detected in the {\em Philae}/COSAC measurements of the 
sub-surface composition of 67P (Goesmann \etal \cite{goesmann15}).
The abundances are presented as a percentage relative to water ice in the 
disk and the data plotted are restricted to within the comet-forming zone 
($\le50$~AU).  
The results show that chemical processing within the disk is able to efficiently 
convert the initial simple ices to more complex species.  
The key process is photodissociation of ice species by UV photons generated by the 
interaction of galactic cosmic rays and stellar X-rays with molecular hydrogen.  
This creates a local source of reactive radicals in the ice mantle to increase complexity.  
Despite a generic model being adopted, the abundances of the four ice species 
presented show remarkable agreement with the ratios derived from the {\em Philae}/COSAC 
measurements (0.3\% for \ce{CH3CN}, 1.5\% for \ce{NH2CHO}, 0.4\% for \ce{HOCH2CHO}, and 
0.3\% for \ce{CH3COCH3}).  
The model results also show good agreement with the observations for \ce{CH3CHO}; however, 
\ce{HNCO} and \ce{CH3NH2} are underpredicted and overpredicted, respectively.  
The reason for this may be because both \ce{HNCO} and \ce{CH3NH2} 
are products formed during successive hydrogenation of OCN and HCN.  
In both cases, the hydrogenation pathways used in the models may be too efficient, 
although the latter has been demonstrated in the laboratory 
(Theul\'{e} \etal \cite{theule11}).    
Conversely, recent laboratory experiments on hydrogenation of HNCO ice 
have shown that hydrogenation does not always lead to saturation 
(Noble \etal \cite{noble15}).
  
The models also do not reproduce the relative abundances of  
\ce{HOCH2CH2COH}, \ce{C2H5CHO}, and \ce{C2H5NH2} observed with {\em Philae}/COSAC 
because the network employed has limited chemistry for these larger 
species.  
Chemical networks, including the one employed here, also do not yet include 
chemistry for \ce{CH3OCN}, neither in the gas phase nor in the ice.    
In light of these exciting results from {\em Philae}/COSAC, the 
expansion of surface networks to better treat these larger species 
should be undertaken including potential pathways to the larger 
COMs mentioned above, which have been demonstrated in the 
laboratory but not yet considered in the models 
(\vg, for \ce{HOCH2CH2COH}; Fedoseev \etal \cite{fedoseev15}).  

\section{Discussion and future outlook}
\label{summary}

The model results show that COMs can be efficiently 
synthesised via surface chemistry in the cold, dense midplanes of 
protoplanetary disks, assuming that the disk inherits simple ices only 
from the parent molecular cloud.  
The abundances attained for several species in the comet-forming 
regions ($\lesssim50$~AU) are on a par with those recently observed on 67P.  
The vital chemical process is cosmic-ray and X-ray-induced photodissociation of 
ice mantle molecules.  
This allows the processing of icy material in the cold dense midplane 
which is otherwise well shielded from both stellar and interstellar UV photons.  
The result for gas-phase \ce{CH3CN} suggest that this species may not be 
an unambiguous tracer of the ice reservoir in the inner regions of disks; 
however, a better understanding of the gas-phase chemistry of \ce{CH3CN} under 
these particular physical conditions is required to confirm this conclusion.  
Gas-phase \ce{CH3OH}, yet to be detected in a protoplanetary disk, 
may be a more robust tracer of the complex ice reservoir.  

The observational results discussed here give a tantalising hint of what 
is to come in the near future.  
The first detection of a complex molecule in a
protoplanetary disk with ALMA gives the community vital information on 
the sensitivities required to detect these heretofore elusive species, that will 
surely be exploited in future cycles.    
In addition, a small fraction only of the data from {\em Rosetta} has been 
published to date, with much more to come.   
Furthermore, a new era of infrared astronomy approaches, 
with MIRI (Mid-InfraRed Instrument) on JWST (James Webb Space Telescope, 
to be launched in 2018) providing unparalleled 
spectral resolution and sensitivity from 5 to 28~$\mu$m 
(Wright \etal \cite{wright04}).  
MIRI is expected to have the sensitivity necessary to observe 
the infrared signature of ice species other than water 
in protoplanetary disks for the first time,  
which may provide the first {\em direct} detection of the 
complex organic reservoir in these objects.  

%%-----------------------------
%%      your bibliography
%%-----------------------------

\end{document}